\documentclass[english,twocolumn]{revtex4}
\usepackage[T1]{fontenc}
\usepackage[latin9]{inputenc}
\usepackage{graphicx}
\usepackage{amssymb}
\usepackage{amsmath}
\usepackage{color}

\newcommand{\bea}{\begin{eqnarray}}

\newcommand{\eea}{\end{eqnarray}}

\newcommand{\beq}{\begin{equation}}

\newcommand{\eeq}{\end{equation}}



\usepackage{babel}

\begin{document}

\title{Magnus expansion approach to  parametric oscillator systems in a thermal bath}

\author{B. Zhu$^{1}$, T. Rexin$^{1}$, and L. Mathey$^{1,2}$}

\affiliation{$^{1}$Zentrum f\"ur Optische Quantentechnologien and Institut f\"ur Laserphysik, Universit\"at Hamburg, 22761 Hamburg, Germany\\
$^{2}$The Hamburg Centre for Ultrafast Imaging, Luruper Chaussee 149, Hamburg 22761, Germany}

\begin{abstract}
We develop a Magnus formalism for periodically driven systems which provides an expansion both in the driving term and the inverse driving frequency,  applicable to isolated  and dissipative systems. 
  We derive explicit formulas for a driving term with a cosine dependence on time, up to fourth order.
  We apply these to the steady state of a classical parametric oscillator coupled to a thermal bath, which we solve numerically for comparison.
  Beyond dynamical stabilization at second order, we find that the higher orders  further renormalize the oscillator frequency, and additionally create a weakly renormalized  effective temperature.
   The renormalized oscillator frequency is quantitatively accurate almost up to the parametric instability, as we confirm numerically.  Additionally, a cut-off dependent term is generated, which indicates the break-down of the hierarchy of time scales of the system, as a precursor to the instability.
   Finally, we apply this formalism to a parametrically driven chain,
    as an example for the control of the dispersion of a many-body system. 
\end{abstract}
\maketitle
\section{Introduction}
The study of periodically driven systems has experienced renewed interest in recent times. Both in solid state  and  ultra cold atom systems, strong periodic driving has been used to control non-equilibrium states. In ultra-cold atom systems, periodic lattice driving has been used to realize an effective, synthetic gauge field, see Ref. \cite{sengstock}. In solid state systems, pump-probe experiments, Ref. \cite{pumpprobeoverview}, on high-T$_{c}$ superconductors and on graphene have been performed, see Refs. \cite{Cavalleri, gierz}. Theoretical studies on light-induced superconductivity were reported in Refs. \cite{hoeppner, okamoto, pumpprobetheory}.

 \begin{figure}
\includegraphics[width=1.08\linewidth]{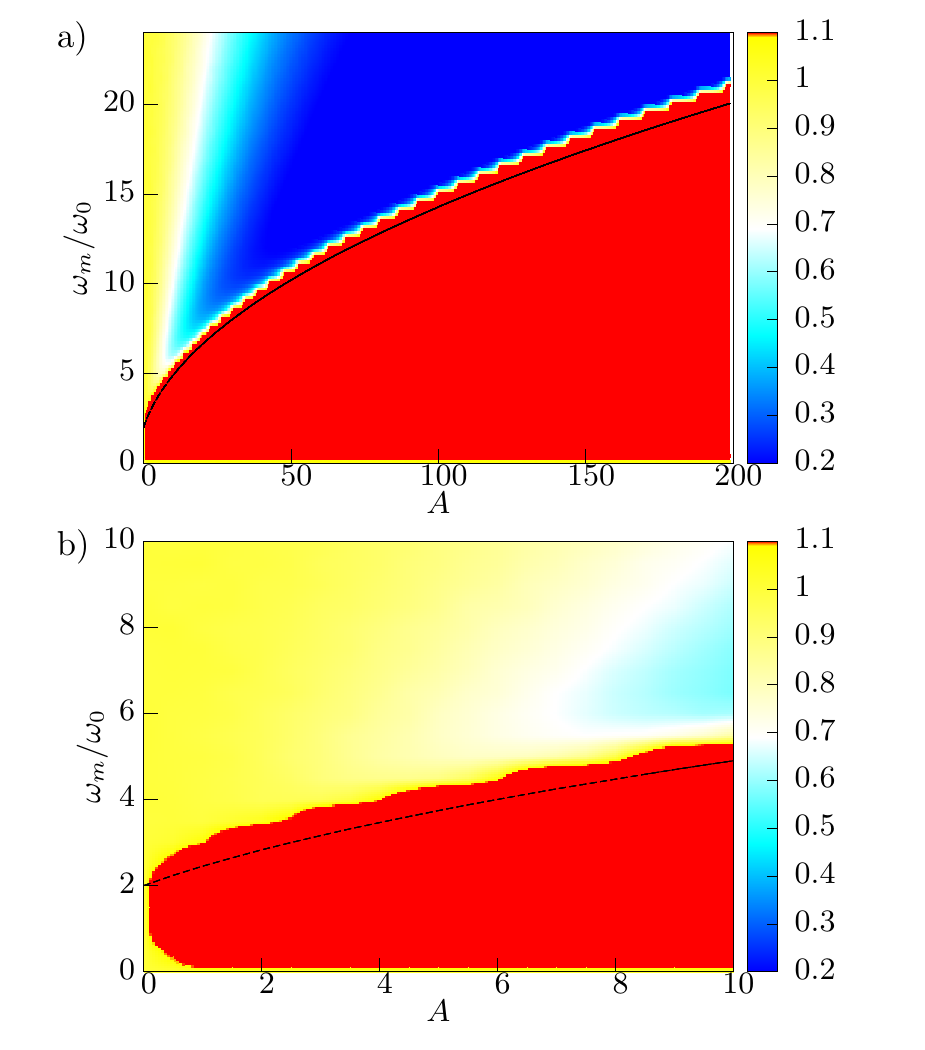}
\caption{We depict the time averaged magnitude of  $\langle x^{2}(t)\rangle/x_{T}^{2}$ in the steady state as a function of driving frequency and driving amplitude.
 Panel (a) and (b) depict the same data on different scales. 
The system displays a power broadened instability emerging from $\sim 2 \omega_{0}$, and a dynamical stabilization for large $\omega_{m}$ and $A$. 
 In panel (a) we show the comparison to Eq. \ref{inst2}, in panel (b) to Eq. \ref{inst1}. }
\label{xsquareAom}
\end{figure}

Remarkably, in both cases,  external high-frequency driving is used to control the low-frequency behavior of each system. The quintessential example for this phenomenon is the  Kapitza effect \cite{Landau}. 
 In the case of the effective synthetic field in an ultra-cold atom system, this process is explicitly described by an approximate, effective low-energy Hamiltonian, which, in contrast to the original, non-driven Hamiltonian, has a synthetic field. In the case of a driven high-T$_{c}$ superconductor, the near-resonant driving of an optical phonon mode results in a modified response in the low-frequency optical conductivity. 
 Both of these observations exemplify the development of a new field of emergence in driven many-body systems.

 In this paper, we give a systematic expansion of the emergent low-energy description of a driven system. This discussion applies and extends the Magnus formalism, as discussed in Refs. \cite{magnus, blanes, salzman, DAlessio}. Our  formalism provides a systematic expansion both in the driving amplitude and the inverse driving frequency, and is applicable to closed and open classical systems, to closed quantum systems. We derive explicit, general expressions for the leading terms beyond second order. 
  As a key example, we apply this formalism to a parametrically driven oscillator, coupled to a thermal bath, Ref. \cite{HaenggiRev}, and determine the properties of its steady state. An insightful discussion of parametric oscillators was given in Refs. \cite{butikov}, as well as in Ref. \cite{Tirapegui}.
    We then apply our  results to a chain of parametrically driven oscillators. This provides insight in how the dispersion of a system can be controlled via parametric driving.
 
This paper is organized as follows: In Sect. \ref{parosc} we describe the  dissipatively coupled, parametrically driven oscillator, and give a discussion of its properties using elementary ansatz functions. In Sect. \ref{ME} we develop the Magnus expansion in full generality first, and then apply it to the parametric oscillator in Sect. \ref{MEparosc}.  
 In Sect. \ref{chain} we discuss the control of the dispersion of a parametrically driven chain of oscillators, and in Sect. \ref{conclusions}, we conclude. 


\section{Parametric oscillator}\label{parosc}
 As  the key example to which we apply the Magnus expansion, we consider a parametrically driven oscillator, described by the Hamiltonian
\bea
H & =&H_{0} + H_{dr}
\eea
with
\bea
H_{0} & =& \frac{p^{2}}{2 m} + \frac{m \omega_{0}^{2}}{2} x^{2}\\
H_{dr} & =& \frac{m \omega_{0}^{2}}{2}  A \cos(\omega_{m} t) x^{2}.
\eea
 $p$ and $x$ are the momentum and spatial coordinate of the oscillator, $m$ is the mass, and $\omega_{0}$ the bare oscillator frequency. $A$ is the amplitude of the parametric driving term, and $\omega_{m}$ is the driving frequency.

We assume that this oscillator is coupled to a thermal bath of temperature $T$, via a dissipative term. 
The resulting equations of motion are of the Langevin form:
\bea
\frac{d x}{d t} &=& \frac{p}{m}\label{Langevinx}\\
\frac{d p}{d t} &=& - m \omega_{0}^{2} (1 + A \cos(\omega_{m} t)) x - \gamma p + \xi.\label{Langevinp}
\eea
$\gamma$ is the damping rate, and $\xi$ describes white noise, with the correlation function $\langle \xi(t_{1}) \xi(t_{2}) \rangle = 2 \gamma k_{B} T m \delta(t_{1} - t_{2})$, where $k_{B}$ is the Boltzmann constant. 
 In thermal equilibrium, in the absence of driving, the system is described by the canonical distribution $\rho_{0}(x, p) = \exp(- \beta H_{0}(x,p))/Z$,  with $\beta = 1/(k_{B} T)$. $Z$ is the partition function, which normalizes this probability distribution.
 For this distribution,  the variances of $x$ and $p$ are $\langle x^{2}\rangle = x_{T}^{2}$ and  $\langle p^{2}\rangle = p_{T}^{2}$, with
 $p_{T} = \sqrt{m k_{B} T}$ and 
 $x_{T} = \sqrt{\frac{k_{B} T}{m \omega_{0}^{2}}}$. 
 Furthermore, we have $\langle x\rangle = \langle p \rangle = \langle x p \rangle =0$. 
   We note that for a classical oscillator, $x_{T}$ and $p_{T}$ can be used to rescale $x$ and $p$. With this choice, the temperature does not appear in any of the remaining quantities, and simply provides an energy scale for the system. For a quantum mechanical oscillator, this rescaling cannot be performed. Here, an additional regime appears in which quantum fluctuations dominate, for $k_{B} T \ll \hbar \omega_{0}$. A full discussion of the driven, dissipative  quantum mechanical oscillator will be given elsewhere. The analysis presented here addresses isolated quantum systems, in addition to dissipative classical systems. 

 \begin{figure}
\includegraphics[width=1.05\linewidth]{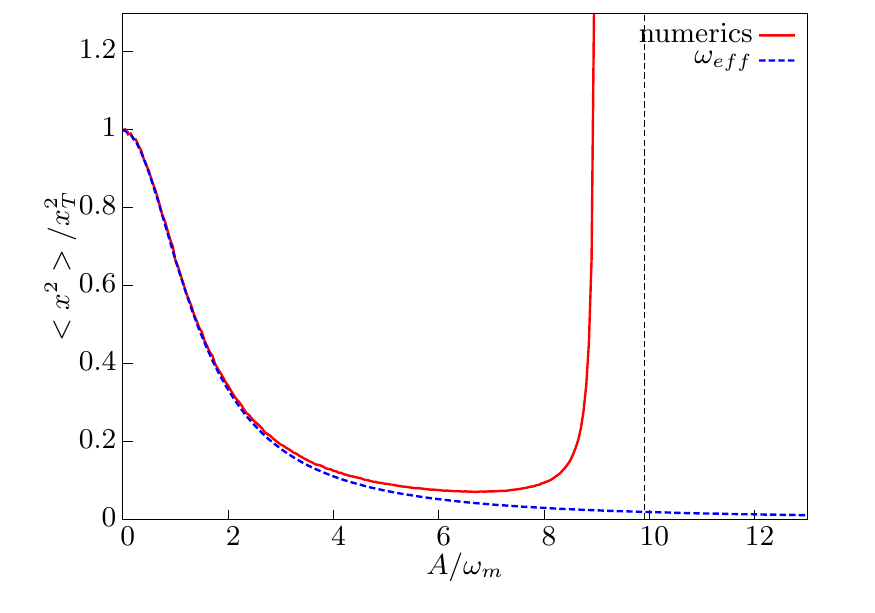}
\caption{We depict the time averaged expectation value of $\langle x^{2}\rangle$, in units of $x_{T}^{2}$, as a function of the driving amplitude $A$, for the driving frequency $\omega_{m}/\omega_{0}=20$,  and for $\gamma/\omega_{0}=0.1$. 
 We compare the numerically obtained result to the prediction in Eq. \ref{elementary2nd}. The dashed, vertical line corresponds to Eq. \ref{inst2}.  }
\label{xsquare}
\end{figure}

    In Fig. \ref{xsquareAom}, we depict the time averaged variance $\langle x(t)^{2}\rangle$, of the steady state of the driven system, as a function of $A$ and $\omega_{m}/\omega_{0}$. Here, and in the examples throughout this paper, we choose  $\gamma/\omega_{0}=0.1$.
 The most striking feature of this plot is the parametric resonance that appears near $\omega_{m}\approx 2 \omega_{0}$, for small $A$. This feature is then power broadened for increasing $A$. In this regime, the magnitude of $\langle x(t)^{2}\rangle$ is increased  by orders of magnitude, compared to the equilibrium value.
  In addition to this strong heating effect, there is a regime for large $\omega_{m}/\omega_{0}$, and large amplitude, for which a reduction of $\langle x(t)^{2}\rangle$ is observed. Here, the parametric driving leads to a dynamic stabilization of the fluctuations of $x$. 
  It is this counterintuitive and quintessential example of reducing fluctuations via high frequency driving that we study systematically in this paper.

 In Fig. \ref{xsquare}, we  show the same quantity in the steady state as a function of $A$, for  a fixed value of  the driving frequency, $\omega_{m}/\omega_{0} =20$, to give a clearer insight into the quantitative behavior. 
  The magnitude of these fluctuations is visibly reduced with increasing driving amplitude. However, eventually this trend of decreasing fluctuations is rapidly reverted, resulting in a steep increase of the fluctuations. As visible from Fig.  \ref{xsquareAom}, this steep increase is due to the power-broadened parametric instability. The onset of this  instability determines the location of the minimal amount of fluctuations that can be achieved with this type of driving. It is therefore imperative to understand the origin of this steep increase of the fluctuations, and provide a systematic approach to determine its behavior.

\begin{figure}
\includegraphics[width=1.05\linewidth]{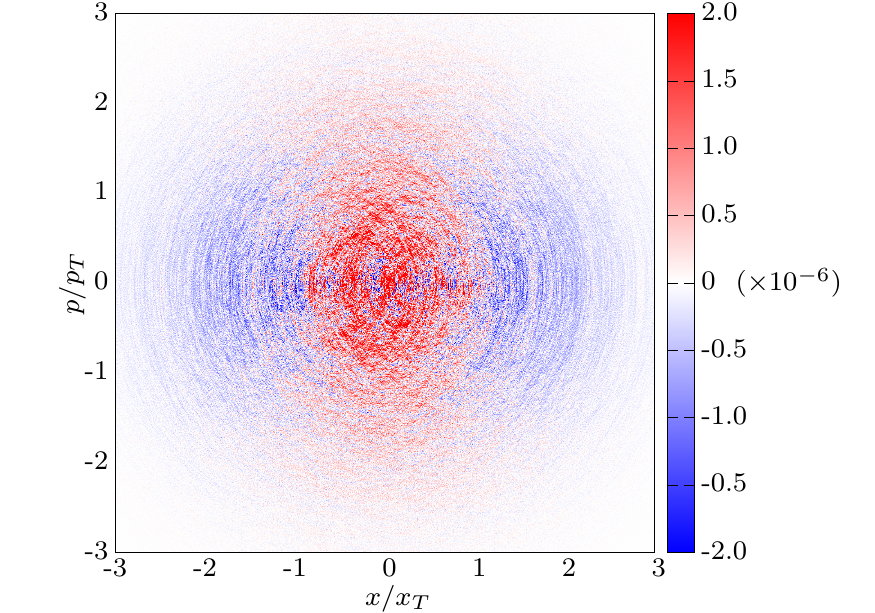}
\caption{Distribution in phase space of the driven system in the steady state. The trajectories of the time evolution have been smoothed out on a time scale of $\sigma = 1/\omega_{0}$. We use  $\omega_{m}/ \omega_{0} = 20$, $A=10$, and $\gamma/\omega_{0} = 0.1$.  The binning size is $\Delta x/x_{T}  =  \Delta p/p_{T} = 0.01$.The reduction of the width of the distribution in the $x$-direction is clearly visible. }
\label{trajectory}
\end{figure}

In Fig. \ref{trajectory} we show a histogram of the distribution $\rho_{dr}$  in the steady state, in comparison to the equilibrium distribution $\rho_{0}$; we show $\rho_{dr} - \rho_{0}$. The distribution $\rho_{dr}$ is generated from trajectories of the Langevin equation, which have been low-frequency filtered via
$x_{c}(t) = \int ds G_{\sigma}(s-t) x(t)$, and similarly for $p_{c}(t)$, derived from $p(t)$. $G_{\sigma}(s)$ is a normalized Gaussian, with a time scale $\sigma$, for which we choose $\sigma = 1/\omega_{0}$. 
In Fig. \ref{trajectory}, furthermore, we choose $A=10$ and $\omega_{m}/\omega_{0}=20$.   
We observe that the width of the distribution along $x$-direction is reduced, due to the dynamical stabilization that is described below. Along the $p$-direction the distribution is only weakly affected. We emphasize that for a quantitative comparison of the driven state to the effective, low-frequency predictions, the exclusion of the high-frequency contributions in the numerics is essential. We elaborate on this point in App. \ref{cutoff}.

%

 \subsection{Elementary approach}\label{elementary}
 Before we develop the renormalization of the oscillator due to the periodic driving systematically in the next section, we give estimates of its behavior by using various ansatz functions. 

We start out by giving an estimate for  the instability regime, and note that a more detailed discussion is given in App. \ref{app_elementary}. 
  We consider the equation of motion of the isolated system, 
  $  \ddot{x} + \omega_{0}^{2} (1 + A \cos(\omega_{m} t)) x =0$. 
  We consider the ansatz
  $x(t) = a_{0} \cos(\omega_{eff} t) + a_{1} \cos((\omega_{m}- \omega_{eff})t)$, 
where $a_{0}$ and $a_{1}$ are constant coefficients.
 We solve for the effective frequency $\omega_{eff}$, which gives
   \bea
 \omega_{eff} &=& \frac{\omega_{m} - \sqrt{\omega_{m}^{2}+ 4 \omega_{0}^{2} - 2 \omega_{0} \sqrt{A^{2} \omega_{0}^{2} + 4 \omega_{m}^{2}} }}{2} 
 \eea
 The instability regime is reached when the expression under the outer square root becomes negative.
  This occurs at
  \bea
  \frac{\omega_{m, pr}}{\omega_{0}} &\approx& \sqrt{2 A+ 4}\label{inst1},
  \eea
  which simplifies to
    \bea
  \frac{\omega_{m, pr}}{\omega_{0}} &\approx&\sqrt{2 A}\label{inst2},
  \eea
  for large $A$.
    This provides an estimate for the instability regime for large driving amplitudes and frequencies, which we show in Fig. \ref{xsquareAom}, and which gives good agreement.

To give an estimate for the renormalization of $\omega_{eff}$, we extend this ansatz to include not only the frequencies $\omega_{eff}$ and $\omega_{m} - \omega_{eff}$, but also  the next three contributing terms, corresponding to the frequencies $\omega_{m} + \omega_{eff}$, $2\omega_{m} - \omega_{eff}$ and  $2\omega_{m} + \omega_{eff}$.
 This ansatz is explicitly written in Eq. \ref{ansatz2}. 
%
 This ansatz  results in Eq. \ref{omegaeff2} for the effective frequency. 
 We solve this equation iteratively in the amplitude $A$, which gives the expansion
\bea
\omega_{eff}^{2} &\approx & \omega_{0}^{2} + \frac{A^{2} \omega_{0}^{4}}{2 (\omega_{m}^{2} - 4 \omega_{0}^{2})} + \frac{25 A^{4} \omega_{0}^{8}}{32 \omega_{m}^{6}}\label{elementary4th}
\eea
At second order in $A$ and at second order in the inverse driving frequency, this is 
\bea
\frac{\omega_{eff}^{2}}{\omega_{0}^{2}} &\approx & 1 + \frac{A^{2} \omega_{0}^{2}}{2 \omega_{m}^{2}}\label{elementary2nd} 
\eea
 This approximation for the effective frequency is shown in Fig. \ref{xsquare}.
  We note that the fourth order term in Eq. \ref{elementary4th} is positive. This is indeed confirmed further down by the systematic Magnus expansion. However, the Magnus expansion determines  the correct prefactor, which differs from the one found here.

\section{Magnus expansion}\label{ME}
We now turn to the Magnus expansion of the system. 
 This expansion provides a time-independent approximation of the low-frequency sector of the system, derived from the original, time-dependent Hamiltonian that describes all frequencies. 
 After deriving general expressions for the Magnus terms beyond second order, 
 we ask the question if and how the key features of the parametric oscillator, the dynamical stabilization and the instability regime, can be captured within this approach.
 We note that these features, as they were described in the previous section, might suggest that such an approach might not be possible in a consistent fashion for the fourth order correction. 
 This is due to the following two observations.
  We observed, as shown in Eq. \ref{elementary4th}, that the fourth order correction has a positive prefactor, which results in an addition stabilization of the oscillator.
  This  term would be derived from a term in an effective Hamiltonian that is of the form $\sim A^4/\omega_{m}^{6}$, with a positive prefactor.
  On the other hand, if the instability of Eq. \ref{inst2} is derived from an effective Hamiltonian, it also needs to be derived from a term of the form  $\sim A^4/\omega_{m}^{6}$,  but now with a negative prefactor. 
  
  Interestingly, as we discuss below, the Magnus expansion provides two types of terms at fourth order. One of them is cut-off independent, and features a positive prefactor. The resulting renormalization due to this term is in agreement with the numerically obtained result.
  The other term is cut-off dependent. It indicates that the hierarchy of time scales that is required for the Magnus expansion breaks down. We interpret this as a precursor of the instability regime, and indeed find that the scaling for this regime, as shown in Eq. \ref{inst2}, is predicted correctly.  
\subsection{Kramers equation}
 To apply the Magnus expansion we formulate the time evolution of the system, Eqs. \ref{Langevinx} and \ref{Langevinp}, 
 as a time evolution of the phase space distribution $\rho(x,p,t)$.
 This is given by the Kramers equation 
\bea
\partial_{t} \rho &=& L(t) \rho.\label{KramersEq}
\eea
 Here, $L(t) = L_{0} + L_{dr}(t)$ 
  with 
\bea
L_{0} \rho  &=& - v \partial_{x}\rho  +\omega_{0}^{2} x \partial_{v}\rho
 + \gamma \Big( \partial_{v}( v \rho) + \frac{k_{B}T}{m} \partial_{vv} \rho\Big)
\eea  
  and
  \bea
  L_{dr} &=& L_{dr, 0} \cos(\omega_{m} t)
  \eea
  with
\bea
L_{dr,0} &=& A \omega_{0}^{2} x \partial_{v}.
\eea
 We refer to Eq. \ref{KramersEq} as the Kramers equation to distinguish it from the Fokker-Planck equation, which we reserve for the over-damped limit, in accordance with the terminology of Ref. \cite{vanKampen}.

\subsection{General expansion}
 We now derive the expansion of the low energy description in full generality. 
 We consider a general, dynamical system that is described by the same equation of motion
\bea
\partial_{t} \rho &=& L(t) \rho\label{generalEoM}
\eea
 as before, without the assumption of the specific form of the equation of motion, as in the previous section. The parametrically driven oscillator will serve as the example to which we apply our results further down. 
 The system under consideration can be either a closed or an open classical system, or a closed quantum system. For a closed quantum system, we interpret the operator $L(t)$ as a Hamiltonian, divided by $i \hbar$, i.e. $L(t) = H(t)/(i \hbar)$. For an open system, we also include dissipative terms, as in Eq. \ref{KramersEq}.
   We again assume that $L(t)$ has the form
 \bea
 L(t) &=& L_{0} + L_{dr}(t),\label{generalLoLd}
 \eea
 where $L_{0}$ describes the time-independent part of the system, and  $L_{dr}(t)$ is the driving term, again of the form
 \bea
  L_{dr}(t) &=& L_{dr, 0} \cos(\omega_{m} t)\label{Ldr}. 
 \eea
 We perform the Magnus expansion in the interaction picture. In this picture, the order of the Magnus expansion coincides with the order of the driving term. In the case of the parametric oscillator, this is the order of the driving amplitude $A$. 
  For the interaction picture we define
\bea
L_{dr,i}(t, s) &=& \exp(- L_{0} s) L_{dr}(t) \exp(L_{0} s)\label{intpic}
\eea  
 where the standard interaction  picture term is $L_{dr, i} (t) = L_{dr,i}(t, t)$.  
 Then the equation of motion is 
\bea
\partial_{t} \rho_{i} &=& L_{dr,i}(t) \rho_{i}.
\eea  
Its solution is
\bea
\rho_{i}(t) &=& T_{t} \exp\Big( \int_{t_{0}}^{t}ds L_{dr, i}(s) \Big) \rho_{i}(t_{0})
\eea
 where $T_{t}$ is the time ordering operator, and $\rho_{i}(t_{0})$ is the initial state at $t_{0}$.
  The Magnus expansion consists of re-expressing this solution in the form
   $\exp(\sum_{i} M_{i})$, where  $M_{i}$ is the Magnus term of $i$-th order, see Ref. \cite{salzman}.

We time average each of these terms over a time interval $[t_{0}, t]$. The time interval is long compared to the driving period, but short compared to the dynamics that is  created by $H_{0}$. For the parametric oscillator, we demand $1/\omega_{0} \gg t - t_{0} \gg 1/\omega_{m}$. The time interval $\Delta t_{c} = t-t_{0}$ is also the inverse of a frequency cut-off $\omega_{c} = 2\pi/\Delta t_{c}$, for which we equivalently  demand $\omega_{0} \ll \omega_{c} \ll \omega_{m}$. 
   For a general system, the frequency $\omega_{0}$ has to be replaced by a typical frequency that is characteristic for the dynamics of $H_{0}$. 
   
   The second order Magnus term in the interaction picture is given by
  \bea
 M_{2,i} &=& - \frac{1}{2}  \int_{t_{0}}^{t} d t_{2} \int_{t_{0}}^{t_{2}} d t_{1}  [L_{dr,i}(t_{1}), L_{dr,i}(t_{2})]\label{M2}
\eea 
 We transfer this expression back to the Schr\"odinger picture, and project this term on the frequency range below $\omega_{c}$. The resulting effective $L_{eff}^{(2)}$ is time-independent, because all the oscillatory contributions oscillate with a frequency above the cut-off frequency.
  The time evolution that results from this term is of the form $\exp(L_{eff}^{(2)} \Delta t_{c})$. Therefore, we can simplify Eq. \ref{M2} by taking the time derivative with respect to $t$, which reduces the number of integrations. 
  The resulting second order term is therefore 
 \bea
L_{eff}^{(2)} &=& \Big[- \frac{1}{2} \int_{t_{0}}^{t} d t_{1}  [L_{dr,i}(t_{1}, \tilde{t}_{1}), L_{dr}(t)]\Big]_{\omega<\omega_{c}}\label{L2}
\eea
 with $\tilde{t}_{1} = t_{1} - t$. 
 We expand the expression in Eq. \ref{intpic} to first order:
\bea
L_{dr,i}(t, s) 
&\approx & L_{dr}(t) - s [L_{0}, L_{dr}(t)]
\eea  
 We use this first order expansion, with $s\rightarrow \tilde{t}_{1}$ and $t \rightarrow t_{1}$, and the time dependence of the driving term,  Eq. \ref{Ldr}, 
 \bea
L_{eff}^{(2)} 
 &\approx&  \frac{1}{2} [[ L_{0}, L_{dr,0}], L_{dr,0}] \nonumber \\
 &&  \times \Big[ \int_{t_{0}}^{t} d t_{1} \tilde{t}_{1}   \cos(\omega_{m}t) \cos(\omega_{m}t_{1})\Big]_{\omega<\omega_{c}}
\eea
The low-frequency part of the time integral, which refers to frequencies below $\omega_{c}$, is
\bea
\Big[ \int_{t_{0}}^{t} d t_{1}  \tilde{t}_{1}  \cos(\omega_{m}t) \cos(\omega_{m}t_{1})\Big]_{\omega<\omega_{c}} &=& \frac{1}{2 \omega_{m}^{2}}
\eea
 Therefore  we obtain 
 \bea
L_{eff}^{(2,2)} 
 &=&   \frac{1}{4 \omega_{m}^{2}} [[ L_{0}, L_{dr,0}], L_{dr,0}]\label{L22}
\eea
 Here, and throughout the paper, we use the notation $L_{eff}^{(n,m)}$ refer to the $n$-th order of the Magnus expansion, and to the $m$-th order in the inverse driving frequency.

\begin{table*}
\begin{tabular}{c | c |c|c|c|c|c|c|c}
    $c_{k_{1}, k_{2}, k_{3}}$ &     $c_{0, 0, 3}$ &     $c_{0, 1, 2}$ & $c_{0, 2, 1}$ & $c_{0, 3, 0}$&   $c_{1, 0, 2}$&  $c_{1, 2, 0}$ & $c_{2, 0, 1}$ & $c_{2, 1, 0}$\\
    \hline 
 & $- \frac{9}{8 \omega_{m}^{6}}$ 
 & $ \frac{7}{16 \omega_{m}^{6}}$&
     $\frac{1}{32 \omega_{m}^{6}} + \frac{\Delta t_{c}^{2}}{4 \omega_{m}^{4}}$
    & $\frac{45}{64 \omega_{m}^{6}}$ 
   & $- \frac{15}{16 \omega_{m}^{6}}$     
    & $\frac{27}{64 \omega_{m}^{6}}$ 
     & $-\frac{33}{32 \omega_{m}^{6}}$ 
      & $\frac{21}{64 \omega_{m}^{6}}$
    \end{tabular}
    \caption{The value of the integrals of the form given in Eq. \ref{cint}, which are necessary to evaluate Eq. \ref{L4} at order $k=3$. }\label{ctable}
\end{table*}

\subsection{Fourth order in $\omega_{m}^{-1}$}
We now derive the next order term in the inverse frequency. 
 We consider the expansion in Eq. \ref{intpic}  to third order
\bea
L_{dr,i}(t, s) 
&\approx & L_{dr}(t) - s [L_{0}, L_{dr}(t)]\nonumber\\
&& + \frac{s^{2}}{2} \text{ad}_{L_{0}}^{2}  L_{dr}(t)  - \frac{s^{3}}{3!} \text{ad}_{L_{0}}^{3}  L_{dr}(t)\label{Ldri}  
\eea  
 where we introduced the notation of the adjoint derivative $\text{ad}_{L_{0}}^{n} L_{dr}(t)$. It is defined via $\text{ad}_{L_{0}}^{n} L_{dr}(t) = [L_{0}, \text{ad}_{L_{0}}^{n-1} L_{dr}(t)]$, and $\text{ad}_{L_{0}}^{0} L_{dr}(t) = L_{dr}(t)$. 
 The term that is quadratic in $s$ gives no low-frequency contribution, therefore $L_{eff}^{(2,3)} =0$.
  The fourth order term is 
 \bea
L_{eff}^{(2, 4)} 
 &=&  \frac{1}{2} [\text{ad}_{L_{0}}^{3}  L_{dr} , L_{dr,0}]\nonumber\\
 && \times \Big[ \int_{t_{0}}^{t} d t_{1}   \frac{\tilde{t}_{1}^{3}}{3!}  \cos(\omega_{m}t) \cos(\omega_{m}t_{1})\Big]_{\omega<\omega_{c}}
\eea
We use the integral property
\bea
\Big[ \int_{t_{0}}^{t} d t_{1} \tilde{t}_{1}^{3} \cos(\omega_{m}t) \cos(\omega_{m}t_{1})\Big]_{\omega<\omega_{c}} &=& - \frac{3}{\omega_{m}^{4}}\nonumber
\eea
which results in 
 \bea
L_{eff}^{(2, 4)} 
 &=& - \frac{1}{4 \omega_{m}^{4}} [\text{ad}_{L_{0}}^{3}  L_{dr} , L_{dr,0}]\label{L24}
\eea
Higher order terms of the form $L_{eff}^{(2, m)}$ can be derived in a similar manner.

\subsection{Fourth order Magnus expansion}
For the fourth order term in the driving term we proceed along the same lines as for the quadratic term in the previous sections.
 The fourth order term in the interaction picture has the form
\bea
&&M_{4, i}\nonumber\\
 &=& - \frac{1}{12}  \int_{t_{0}}^{t}  d t_{4}\int_{t_{0}}^{t_{4}} d t_{3}  \int_{t_{0}}^{t_{3}} d t_{2} \int_{t_{0}}^{t_{2}} d t_{1}\nonumber\\
&& \Big(  [L_{dr,i}(t_{1}), [[L_{dr,i}(t_{2}),L_{dr,i}(t_{3})],L_{dr,i}(t_{4})]]\nonumber\\
&& +[[L_{dr,i}(t_{1}), [L_{dr,i}(t_{2}),L_{dr,i}(t_{3})]],L_{dr,i}(t_{4})]\nonumber\\
&& + [[L_{dr,i}(t_{1}), L_{dr,i}(t_{2})],[L_{dr,i}(t_{3}),L_{dr,i}(t_{4})]]\nonumber\\
&& +[[L_{dr,i}(t_{1}), L_{dr,i}(t_{3})],[L_{dr,i}(t_{2}),L_{dr,i}(t_{4})]] \Big) 
\eea
 Again, we transform this expression to the Schr\"odinger picture.
  We project this term on the low-frequency regime. 
 Interestingly, we find two contributions, as we show below.
 The first is proportional to $\Delta t_{c}$. Therefore it lends itself to an interpretation as an effective low-energy description.
  The second term is cubic in $\Delta t_{c}$, which means that we can write 
  \bea
  \Big[M_{4}\Big]_{\omega<\omega_{c}} & =& L_{eff}^{(4)} \Delta t_{c} + \tilde{L}_{eff, c}^{(4)} \Delta t_{c}^{3},\label{M4lf}
  \eea
  and we also introduce the definition  $L_{eff, c}^{(4)} =  \tilde{L}_{eff, c}^{(4)} \Delta t_{c}^{2}$.
  We again obtain the operators $L_{eff}^{(4)}$ and $L_{eff, c}^{(4)}$ by considering the low-frequency sector  of the time derivative of $M_{4}$, i.e.
     \begin{widetext}
 \bea
   L_{eff}^{(4)} + 3 L_{eff, c}^{(4)}
&=&  \Big[ - \frac{1}{12}  \int_{t_{0}}^{t} d t_{3}  \int_{t_{0}}^{t_{3}} d t_{2} \int_{t_{0}}^{t_{2}} d t_{1}
 \Big(  [L_{dr,i}(t_{1}, \tilde{t}_{1}), [[L_{dr,i}(t_{2}, \tilde{t}_{2}),L_{dr,i}(t_{3}, \tilde{t}_{3})],L_{dr}(t)]]\nonumber\\
&& +[[L_{dr,i}(t_{1}, \tilde{t}_{1}), [L_{dr,i}(t_{2}, \tilde{t}_{2}),L_{dr,i}(t_{3}, \tilde{t}_{3})]],L_{dr}(t)]
 + [[L_{dr,i}(t_{1}, \tilde{t}_{1}), L_{dr,i}(t_{2}, \tilde{t}_{2})],[L_{dr,i}(t_{3}, \tilde{t}_{3}),L_{dr}(t)]]\nonumber\\
&& +[[L_{dr,i}(t_{1}, \tilde{t}_{1}), L_{dr,i}(t_{3}, \tilde{t}_{3})],[L_{dr,i}(t_{2}, \tilde{t}_{2}),L_{dr}(t)]] \Big)\Big]_{\omega<\omega_{c}}\label{L4} 
 \eea
    \end{widetext}
 with $\tilde{t}_{i} = t_{i} -t$.     
  The factor of $3$ in front of $L_{eff, c}^{(4)}$ is due to the derivative of Eq. \ref{M4lf}. 
 We use the expansion of $L_{dr,i}$, given in Eq. \ref{Ldri}. We order the resulting terms  according to the combined order of the times $\tilde{t}_{i}$, i.e. $\tilde{t}_{1}^{k_{1}}\tilde{t}_{2}^{k_{2}}\tilde{t}_{3}^{k_{3}}$, and $k=k_{1} + k_{2} +k_{3}$. 
  The first and second order terms with $k=1$ and $k=2$ gives no contribution.
  For the $k=3$ term, a number of contributions are generated in this expansion.
   These contain time integrals of the form
  \bea
  c_{k_{1}, k_{2}, k_{3}}&=& \Big[ \cos(\omega_{m} t) \int_{t_{0}}^{t} d t_{3} \int_{t_{0}}^{t_{3}} d t_{2} \int_{t_{0}}^{t_{2}} d t_{1}  \cos(\omega_{m} t_{1})\nonumber\\
&& \cos(\omega_{m} t_{2})  \cos(\omega_{m} t_{3})  \tilde{t}_{1}^{k_{1}} \tilde{t}_{2}^{k_{2}} \tilde{t}_{3}^{k_{3}}\Big]_{\omega<\omega_{c}}\label{cint} 
  \eea
The integrals that are necessary to derive $L_{eff}^{(4)} + 3 L_{eff, c}^{(4)}$ are given in Table \ref{ctable}.  
All the terms that scale as $1/\omega_{m}^{6}$ contribute to $L_{eff}^{(4)}$.
 These are written out and simplified in App. \ref{ME4}.
  We obtain $L_{eff}^{(4)}$ to be
 \bea
&&    L_{eff}^{(4,6)}\nonumber\\
 &= &\frac{1}{12 \omega_{m}^{6}} \Big( \frac{39}{64} [L_{dr,0}, [[  \text{ad}_{L_{0}}^{3} L_{dr,0}  ,L_{dr,0}],L_{dr,0}]]\nonumber\\
 &  & + \frac{61}{64}  [L_{dr,0}, [[  \text{ad}_{L_{0}}^{2} L_{dr,0}, \text{ad}_{L_{0}} L_{dr,0}],L_{dr,0}]]\nonumber\\   
    & & + \frac{87}{32} [[  \text{ad}_{L_{0}}^{2} L_{dr,0}, [    \text{ad}_{L_{0}} L_{dr,0}, L_{dr,0} ]],L_{dr,0}]\nonumber\\
&  &- \frac{3}{32} [[ \text{ad}_{L_{0}}^{2} L_{dr,0},L_{dr,0}] , [ \text{ad}_{L_{0}} L_{dr,0} ,L_{dr,0}]]\Big)\label{L46}
 \eea 
  The term that scales as $\Delta t_{c}^{2}/\omega_{m}^{4}$, which is due to the $c_{0,2,1}$ integral, gives $3 L_{eff, c}^{(4)}$.
 Therefore,  
  the cut-off dependent contribution is
   \bea
 &&   L_{eff, c}^{(4,6)}\nonumber\\
&=&    \frac{\Delta t_{c}^{2}}{144 \omega_{m}^{4}}  [L_{dr,0}, [[  \text{ad}_{L_{0}}^{2} L_{dr,0}, \text{ad}_{L_{0}} L_{dr,0}],L_{dr,0}]]\label{L46c}
 \eea 
  We emphasize again, that for any system that can be written in the form of Eqs. \ref{generalEoM}, \ref{generalLoLd}, and  \ref{Ldr}, the results given in Eqs. \ref{L22}, \ref{L24}, \ref{L46}, and  \ref{L46c} apply.
  They constitute the main conceptual result of this paper.


%

\begin{figure}
\includegraphics[width=1.1\linewidth]{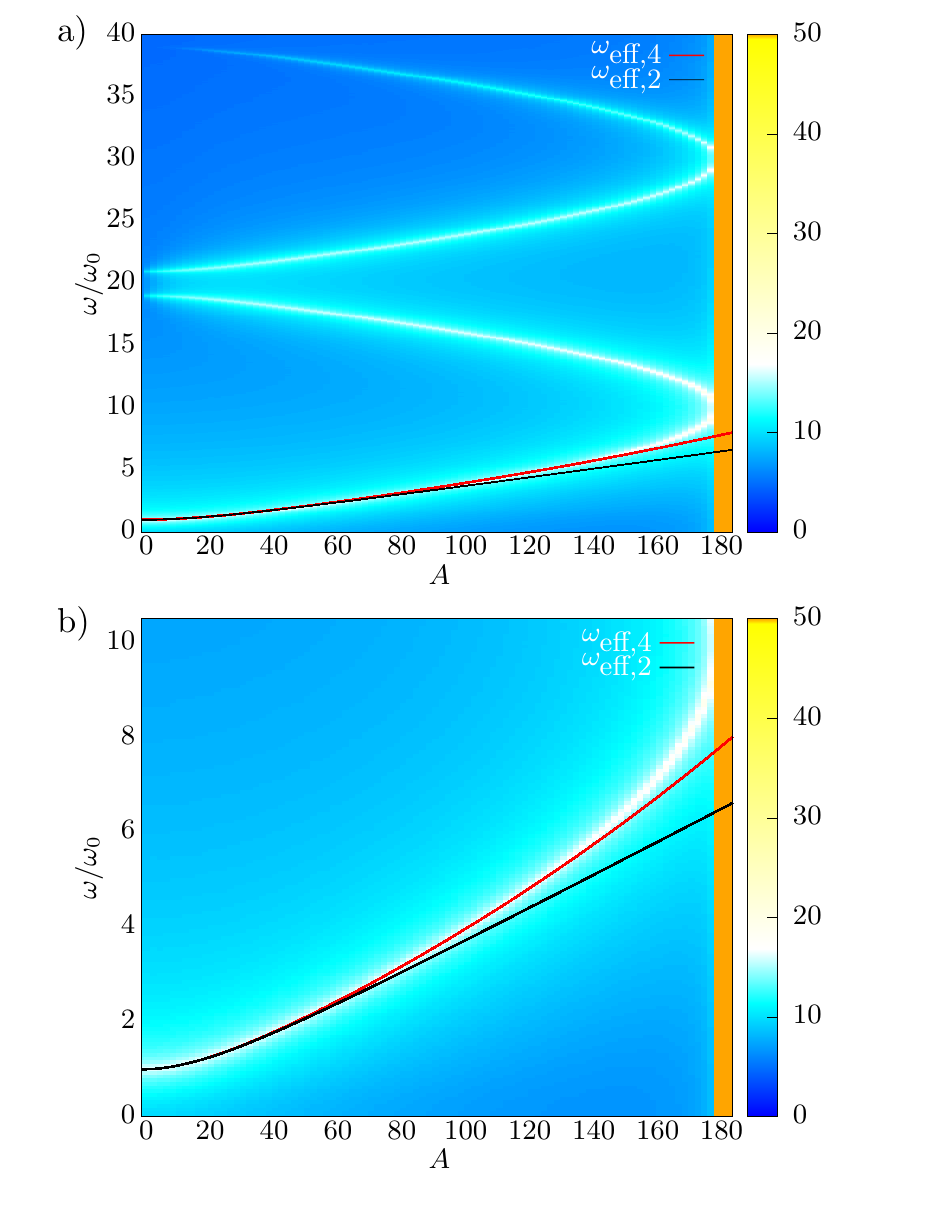}
\caption{Power spectrum $S_{p}(\omega)$ as a function of the driving amplitude $A$, depicted on a logarithmic scale. For the driving frequency we use $\omega_{m}/\omega_{0} = 20$.
 We show the second order estimate of the effective frequency, $\omega_{eff,2}$, which refers to Eq. \ref{oeff2}. Additionally, we show the fourth order estimate  $\omega_{eff,4}$, based on Eq. \ref{oeff4}.  }
\label{psd}
\end{figure}

\section{Magnus expansion of the parametric oscillator}\label{MEparosc}
 We now apply our results to the case of  the parametric oscillator, introduced above.
 For the $L_{eff}^{(2,2)}$ correction we use Eq. \ref{L22}, and find
 \bea
 L_{eff}^{(2,2)} &=&   \frac{A^{2} \omega_{0}^{4}}{2 \omega_{m}^{2}} x \partial_{v}\label{paroscL22}
 \eea
This implies a renormalization of the oscillator frequency of the form
\bea
 \frac{\omega_{eff}^{2}}{\omega_{0}^{2}} &=& 1 + \frac{A^{2} \omega_{0}^{2}}{2 \omega_{m}^{2}}\label{oeff2} 
\eea
  This coincides with the second order term that was obtained in Eq. \ref{elementary2nd}.  
  At the fourth in the inverse driving frequency we have
   \bea
 L_{eff}^{2,4}
&=& \frac{A^{2} \omega_{0}^{4}}{4 \omega_{m}^{4}} \Big( 2 (4 \omega_{0}^{2} - \gamma^{2}) x \partial_{v}
 + 8 \gamma (T/m) \partial_{vv} \Big)
 \eea 
  where we applied Eq. \ref{L24}. 
   Interestingly, in addition to a further renormalization of the oscillator frequency, a renormalization of the temperature is created:
 \bea
 \frac{\omega_{eff}^{2}}{\omega_{0}^{2}} &=& 1 + \frac{A^{2} \omega_{0}^{2}}{2 \omega_{m}^{2}} + \frac{A^{2} \omega_{0}^{2} (4 \omega_{0}^{2} - \gamma^{2}) }{2 \omega_{m}^{4}}\\
 \frac{T_{eff}}{T} &=& 1 + \frac{2 A^{2}\omega_{0}^{4}}{\omega_{m}^{4}}\label{Teff}.
 \eea
 It is generated because the white-noise dissipative term contains  fluctuations at all frequencies, in particular at the driving frequency $\omega_{m}$. 
  This results in an additional renormalization of the low-frequency regime, via time averaging, of the system at this higher order. For the example presented here, the magnitude of the renormalization is small. However, non-linear systems will in general create non-linear effective dissipative terms at this order. 
 Finally we determine the two terms at order $A^{4}$.
  The cut-off independent term is
 \bea
  L_{eff}^{(4,6)} &=& \frac{107}{96} \frac{A^{4} \omega_{0}^{8}}{\omega_{m}^{6}} x \partial_{v}
 \eea
 This term generates an additional renormalization of the oscillator frequency, resulting in
  \bea
 \frac{\omega_{eff}^{2}}{\omega_{0}^{2}} &=& 1 + \frac{A^{2} \omega_{0}^{2}}{2 \omega_{m}^{2}} + \frac{A^{2} \omega_{0}^{2} (4 \omega_{0}^{2} - \gamma^{2}) }{2 \omega_{m}^{4}} \nonumber\\
 && + \frac{107 A^{4} \omega_{0}^{6}}{96 \omega_{m}^{6}}\label{oeff4}
 \eea
 We note that this renormalization at fourth order in $A$ has a positive prefactor, as in the estimate in Eq. \ref{elementary4th}. However, the systematic Magnus expansion gives the correct magnitude of the  prefactor. 
 
 In Fig. \ref{psd} we depict the power spectrum $S_{p}(\omega)$ of the momentum $p$ in steady state, 
   as a function of the driving amplitude $A$, and for the fixed driving frequency $\omega_{m}/\omega_{0}=20$. The power spectrum is defined via
   \bea
   S_{p} (\omega) &=& \langle p(-\omega) p(\omega) \rangle
   \eea
  with $p(\omega) = (1/\sqrt{T}_{s}) \int dt' \exp(- i \omega t') p(t')$, where $T_{s}$ is the sampling interval.
  At $A=0$ the power spectrum reduces to that of a harmonic oscillator, with a single peak at $\omega_{0}$. As the driving is turned on, additional peaks appear at $n \omega_{m} \pm \omega_{0}$, where $n$ is an integer describing the Floquet band. We note that these frequencies are approximately the ones that were used in the ansatz functions in Sect. \ref{elementary} and App. \ref{app_elementary}. 
  With increasing driving amplitude, the effective oscillator frequency increases. We compare this increase to the 
  second order prediction, given in Eq. \ref{oeff2}, and the fourth order prediction, Eq. \ref{oeff4}.
   The fourth order estimate describes the oscillator frequency well almost up to the instability, which is reached around $A\approx 180$, in this example.
  We emphasize that the orange bar at $A\gtrsim 180$ is numerical data. Here, the magnitude of power spectrum increases rapidly by many orders of magnitude. 
 
  The cut-off dependent term is
 \bea
  L_{eff, c}^{(4,6)} &=& - \frac{ A^{4} \omega_{0}^{8} \Delta t_{c}^{2}}{18 \omega_{m}^{4}}  x \partial_{v}\label{paroscL46c}
 \eea
This term competes with the previously discussed terms which stabilize the oscillator.
 For simplicity we only consider the dominant term of the effective frequency,  Eq. \ref{paroscL22}. We relate the time scale $\Delta t_{c}$
 to a frequency cut-off via $\Delta t_{c} = 2 \pi/\omega_{c}$.
 We assume to be in the strongly renormalized regime, $\omega_{eff}^{2}/\omega_{0}^{2} \approx \frac{A^{2} \omega_{0}^{2}}{2 \omega_{m}^{2}}$.
  Therefore, $L_{eff, c}^{(4,6)}$ competes with this renormalization if
  \bea
  \frac{A^{2} \omega_{0}^{2}}{2 \omega_{m}^{2}} &\approx& \frac{2 \pi^{2} A^{4} \omega_{0}^{6}}{9 \omega_{m}^{4} \omega_{c}^{2}}
  \eea
 This results in the criterium
\bea
\frac{\sqrt{\omega_{m} \omega_{c}}}{\omega_{0}} &\approx & \sqrt{A}\label{omcA} 
\eea
If we consider a cut-off frequency chosen as fraction of the driving frequency, and therefore $\omega_{c} \sim \omega_{m}$, we recover 
\bea
\frac{\omega_{m}}{\omega_{0}} &\approx & \sqrt{A}
\eea
which displays the same scaling as in Eq. \ref{inst2}.
 The scaling displayed in Eq. \ref{omcA} can also be motivated by comparing the cut-off frequency $\omega_{c}$ to $\omega_{eff}/\omega_{0}\sim A \omega_{0}/\omega_{m}$.
  Again, this condition indicates that the originally assumed hierarchy of energy scales is no longer valid. This property of the system derives from the cut-off dependent term $L_{eff, c}^{(4,6)}$. While this term in itself cannot be interpreted as a contribution to the effective equation of motion, it can give an insight into the breakdown of the necessary hierarchy of time scales of the system.

\section{Parametrically driven chain}\label{chain}   
 We apply this formalism to the stabilization of a chain of oscillators via parametric driving.
  This, and related mechanisms, have been considered in the context of light enhanced superconductivity, with the following motivation. 
  If we imagine a complex order parameter field describing  fluctuating superconducting order, a key feature of this system is its phase stiffness. In equilibrium, it controls the superconducting stability and the critical current. The phase stiffness in turn is related to how steeply the dispersion of the system increases with increasing momentum.
    Therefore, one possible explanation of light enhanced superconductivity might entail stabilizing and steepening the dispersion of the system. 
  
 We here give the simplest, yet generic, case of a one-dimensional chain of oscillators. 
 The system is described by $H=H_{0}+H_{dr}(t)$, with
\bea
H_{0} & =& \sum_{i}\Big( \frac{p_{i}^{2}}{2 m} + \frac{m \omega_{0}^{2}}{2} (x_{i} - x_{i+1})^{2}\Big)
\eea
 with $i = 1, \ldots, N$.
The driving term is
\bea
H_{dr} & =& \sum_{i}\frac{m \omega_{0}^{2}}{2}  A \cos(\omega_{m} t) (x_{i}-x_{i+1})^{2}
\eea
 We therefore have a parametrically driven lattice of oscillators.
    We Fourier transform the system via 
   $x_{i} = \frac{1}{\sqrt{N}} \sum_{k} \exp(i k r_{i}) x_{k}$ and $p_{i} = \frac{1}{\sqrt{N}} \sum_{k} \exp(i k r_{i}) p_{k}$, and note that $[x_{k_{1}}, p_{k_{2}}] = i \hbar \delta_{k_{1}, - k_{2}}$.
 The Langevin equations for the system are
\bea
\frac{d x_{k}}{d t} &=& \frac{p_{k}}{m}\label{chainx}\\
\frac{d p_{k}}{d t} &=& - m \omega_{k, 0}^{2} (1 + A \cos(\omega_{m} t))  x_{k}\label{chainp} 
 - \gamma p_{k} + \xi_{k}
\eea
 with $\langle \xi_{k_{1}}(t_{1}) \xi_{k_{2}}(t_{2}) \rangle = 2 \gamma k_{B} T m \delta_{k_{1}, -k_{2}} \delta(t_{1} - t_{2})$.
 In real space this corresponds to  $\langle \xi_{i}(t_{1}) \xi_{j}(t_{2}) \rangle = 2 \gamma k_{B} T m \delta_{ij} \delta(t_{1} - t_{2})$.
           The dispersion $\omega_{k, 0}$  is
 \bea
 \omega_{k, 0} &=& \omega_{0} \sqrt{2 - 2 \cos k} = 2 \omega_{0} |\sin (k/2)| \label{chain_k0}
 \eea
   We observe that the equations \ref{chainx} and \ref{chainp} are equivalent to Eqs. \ref{Langevinx} and \ref{Langevinp}, with the replacement $\omega_{0} \rightarrow \omega_{k,0}$.  
     We can therefore apply the results for the single oscillator, Eq. \ref{oeff4}, to each momentum mode and obtain the effective dispersion
    \bea
 \omega_{k, eff}^{2} &=& \omega_{k, 0}^{2} \Big(1 + \frac{A^{2} \omega_{k, 0}^{2}}{2 \omega_{m}^{2}} + \frac{A^{2} \omega_{k, 0}^{2} (4 \omega_{k, 0}^{2} - \gamma^{2}) }{2 \omega_{m}^{4}} \nonumber\\
 && + \frac{107 A^{4} \omega_{k, 0}^{6}}{96 \omega_{m}^{6}}\Big)\label{okeff4}
 \eea
 The second order correction, derived from $L_{eff}^{(2,2)}$, contains contributions of the form $\sim \cos 2k$. This can be seen by substituting $\omega_{k,0} = 2\omega_0|\sin (k/2)|$, see Eq.(\ref{chain_k0})). This describes coupling to the next-nearest neighbor, induced by the periodic driving, because a next-nearest coupling term of the form
 $\sum_{i} x_{i} x_{i+2}$ gives rise to $\cos 2k$ terms in momentum space when Fourier transformed. 
  When substituting  Eq.(\ref{chain_k0}), in the term that is quadratic in $A$ and quartic in $\omega_{m}^{-1}$, we obtain  terms up to $\sim \cos 3k$, which corresponds to coupling to the third neighbor. Finally the term quartic in $A$ contains coupling to the fourth nearest neighbor.
 
 \begin{figure}
\includegraphics[width=1.1\linewidth]{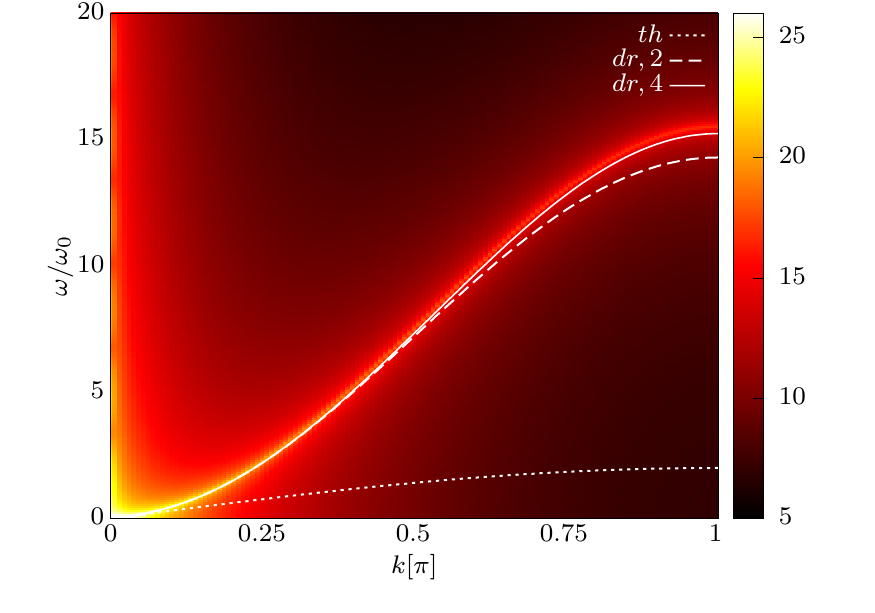}
\caption{two point correlation function $ \mathcal{G}(k,\omega)$, depicted on a logarithmic scale. We use $\omega_{m}/\omega_{0} = 80$ for the driving frequency with driving amplitude $ A = 400 $.
 We compare the numerics with the analytical estimates(white lines), respectively thermal situation(dotted line), 2nd order estimate(dashed line) and 4th order estimate(solid line) of the effective frequency, $\omega_{k,eff}$, which refers to Eq. \ref{okeff4}. }
\label{psd_lattice}
\end{figure}
 
 In Fig. \ref{psd_lattice} we show the two point correlation $\mathcal{G}(k,\omega)$ in momentum and coordinate space in the steady state for a one-dimension chain of parametrically driven oscillators. The two point correlation function is defined by 
 \bea
  	\mathcal{G}(k,\omega) = < X(-k,-\omega) X(k,\omega)> 
 \eea
 with 
 \bea
 	 X(k,\omega) = \frac{1}{\sqrt{K_s T_s}} \int dt'\int dr' x(r',t') \exp(-ikr') \exp(-i\omega t') \nonumber
 \eea
 where $ T_s $ and $K_s$ are sampling time interval and space interval, respectively.
 Compared with the nondriven situation, the driven dispersion line has a steeper slope, which means the driving term stiffens the system significantly. We compare the numerics with effective dispersion $\omega_{k,eff}$, Eq. \ref{okeff4}. It is clearly seen that at higher k modes, the 2nd order correction deviates from the numerics while the 4th order correction describes the numerics precisely.

 We also observe that the strongest renormalization of the dispersion occurs at its upper edge. This includes the onset of the parametric instability. We now have the condition 
\bea
\frac{\omega_{m}}{\omega_{k, 0}} &\approx & \sqrt{A}
\eea  
 which is first reached for the maximum of  the band. This sets the upper limit for the driving amplitude $A$ that can be used to stabilize the dispersion.
  However, we note that, depending on the physical system, the range of $A$ might be much more limited. For example, the value of the spring constant between neighboring oscillators might not allow for negative values, meaning that $A<1$. With this constraint, the magnitude of the renormalization is small, of the order of $\omega_{k, 0}^{2}/ \omega_{m}^{2}$.

%

\section{Conclusions}\label{conclusions}
 We have developed a systematic Magnus expansion in the driving term and the inverse driving frequency. In this formalism we have derived explicit expressions
  for a system with a driving term with cosine time dependence. This system can be either a quantum mechanical system, or a classical system including dissipative terms.
   The main, conceptual formulas are given in Eqs. \ref{L24}, \ref{L46} and \ref{L46c}, which are the terms beyond the widely discussed lowest order term in Eq. \ref{L22}. 
 At fourth order in the driving term, we find two contributions, one cut-off independent and one cut-off dependent. 
 The cut-off independent term contributes to the effective Kramers or Hamilton operator, whereas the increasing magnitude of the cut-off dependent term indicates  the 
 breakdown of the hierarchy of time scales that was originally assumed.
  We apply this formalism to a parametrically driven oscillator, coupled to a thermal bath, and to a parametric oscillator chain.
  We obtain the magnitude of stabilization that can be achieved for these systems, and the onset of the instability.
  We emphasize that our formalism can be applied to a wide range of driven systems, including non-linear systems and many-body systems.
  It will be of particular interest to the emerging field of controlling many-body systems via external driving.

\begin{acknowledgments}
 We gratefully acknowledge discussions with Andrea Cavalleri, Robert H\"oppner and Junichi Okamoto.
 We acknowledge support from the Deutsche Forschungsgemeinschaft through the SFB 925 and through Project No. MA 5900/1-1,  the Hamburg Centre for Ultrafast Imaging, and from the Landesexzellenzinitiative Hamburg, supported by the Joachim Herz Stiftung. B.Z. acknowledges support from the China Scholarship Council, under scholarship No. 2012 0614 0012. 

\end{acknowledgments}

\appendix

\section{Frequency cut-off}\label{cutoff}  
  In this section we discuss the  comparison of the predictions of the effective description to the observables extracted from the full system.
  Because the effective description is a low-frequency description, it is, in general, imperative to apply a frequency cut-off on the observables, for a quantitative comparison.
   While for some observables depend only weakly on the introduction of this cut-off, in general the low-pass filtered observable will differ from the observable that includes all frequencies. 
  
\begin{figure}
\includegraphics[width=1.0\linewidth]{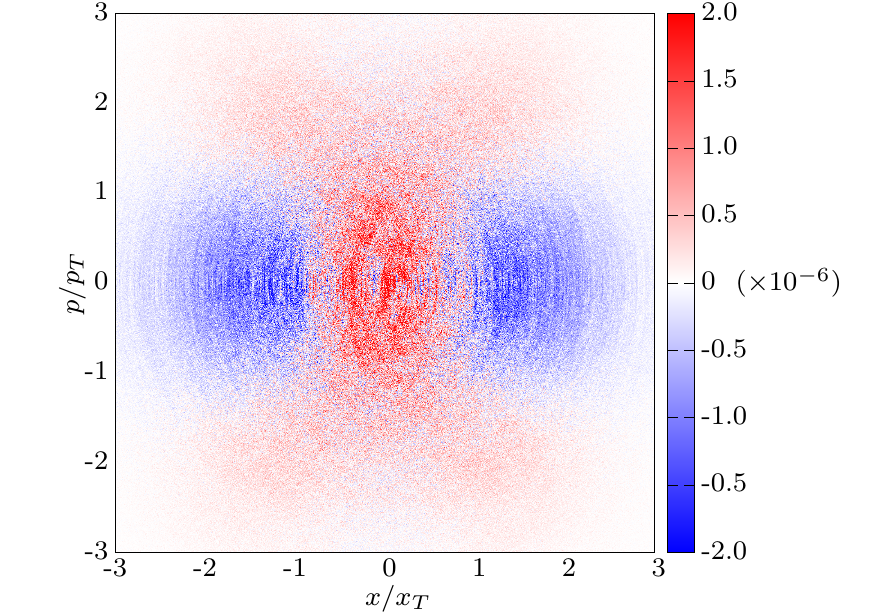}
\caption{Distribution in phase space of the driven system in the steady state, for the same parameters as in Fig. \ref{trajectory}, but without the low-frequency filtering.  
 For this distribution, an increase of the width in the $p$-direction is observed, which is due to high-frequency contributions.}
\label{histogramcutoff}
\end{figure}

 As discussed in Sect. \ref{parosc},   we have depicted the phase space distribution that is derived from the low-frequency filtered trajectories $(x_{c}(t), p_{c}(t))$  in Fig. \ref{trajectory}.
   For comparison, 
  we depict the phase space distribution that is derived from original trajectories $(x(t), p(t))$ that include all frequencies,  in Fig. \ref{histogramcutoff}. 
    As is clearly visible, for this distribution a broadening of the distribution in the $p$-direction occurs, in contrast to Fig. \ref{trajectory}.
%
%
%

\begin{figure}
\includegraphics[width=1.0\linewidth]{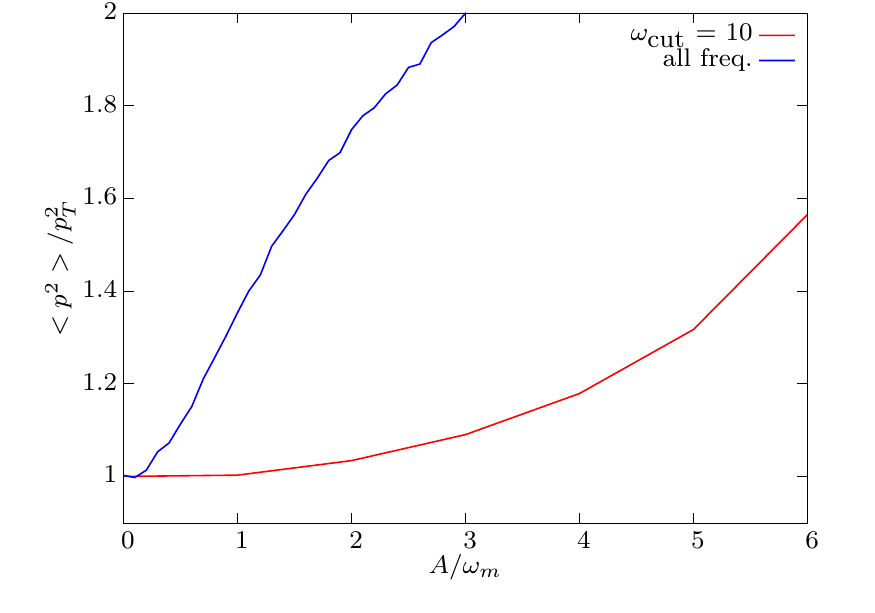}
\caption{Time average value of $p^{2}(t)$, shown as the blue line, and $p_{c}^{2}(t)$, shown as the red line. The additional increase of $p^{2}(t)$ is due to high-frequency contributions. For a quantitative comparison to effective, low-frequency descriptions, the low-frequency filtered  observable has to be used. }
\label{varpcutoff}
\end{figure}

   To elaborate on this further, we depict the time average of $\langle p_{c}(t)^{2}\rangle$ and $\langle p(t)^{2}\rangle$ in the steady state, as a function of $A$, in Fig. \ref{varpcutoff}.
   $\langle p(t)^{2}\rangle$ has a strong dependence on $A$, which is approximately quadratic.
   $\langle p_{c}(t)^{2}\rangle$, however, has only a very weak $A$ dependence, only given the weak temperature renormalization  that a was given in Eq. \ref{Teff}.
%
%
   
\section{Elementary ansatz}\label{app_elementary}
We consider the equation of motion for the isolated system 
  \bea
  \ddot{x} + \omega_{0}^{2} (1 + A \cos(\omega_{m} t)) x &=&0.
  \eea 
  To estimate the regime in which the instability of the system occurs, 
  we consider the ansatz
 \bea
 x(t) &=& a_{0} \cos(\omega_{eff} t) + a_{1} \cos((\omega_{m}- \omega_{eff})t)
 \eea
where $a_{0}$ and $a_{1}$ are constant coefficients. $\omega_{eff}$ is the effective oscillation frequency, which we solve for. 
  Substituting this ansatz in the equation of motion, and ignoring further frequencies, this results in the equations
 $(\omega_{eff}^{2} - \omega_{0}^{2})a_{0} = A \omega_{0}^{2} a_{1}/2$ 
 and  $((\omega_{m} - \omega_{eff})^{2} - \omega_{0}^{2})a_{1} = A \omega_{0}^{2} a_{0}/2$.
  We eliminate $a_{0}$ and $a_{1}$ and obtain the equation
  \bea
 ( \omega_{eff}^{2} - \omega_{0}^{2}) ( (\omega_{m}-\omega_{eff})^{2} - \omega_{0}^{2}) &=& \frac{A^{2}\omega_{0}^{4}}{4} 
  \eea 
 The resulting $\omega_{eff}$ is
  \bea
 \omega_{eff} &=& \frac{\omega_{m} - \sqrt{\omega_{m}^{2}+ 4 \omega_{0}^{2} - 2 \omega_{0} \sqrt{A^{2} \omega_{0}^{2} + 4 \omega_{m}^{2}} }}{2} 
 \eea
 The parametric resonance is reached when the expression under the square root becomes negative.
  We note that the effective frequency $\omega_{eff}$  increases monotonously, with increasing $A$. The instability occurs when the two frequencies $\omega_{eff}$ and $\omega_{m} - \omega_{eff}$ equal each other. We confirm this behavior by calculating the power spectrum of the driven state, which is shown in Fig. \ref{psd}.


 To give a more accurate estimate of the renormalization of $\omega_{eff}$, we 
 consider following ansatz 
 \bea
 x(t) &=& a_{0} \cos(\omega_{eff} t) + a_{1} \cos((\omega_{m}- \omega_{eff})t)\nonumber\\
 && + a_{2} \cos((\omega_{m}+ \omega_{eff})t) + a_{3} \cos((2\omega_{m}- \omega_{eff})t)\nonumber\\
 && + a_{4} \cos((2 \omega_{m}+ \omega_{eff})t)\label{ansatz2}.
 \eea
 When we substitute this in the equation of motion, we obtain the following equation for $\omega_{eff}$: 
\bea
&&\omega_{eff}^{2} - \omega_{0}^{2}\nonumber\\
 &=& - \frac{A^{2} \omega_{0}^{4}}{4} \Big( \frac{1}{\omega_{0}^{2} - (\omega_{m}-\omega_{eff})^{2} +\frac{A^{2} \omega_{0}^{4}/4}{(2\omega_{m} - \omega_{eff})^{2} - \omega_{0}^{2}} }\nonumber\\
&&  +\frac{1}{\omega_{0}^{2} - (\omega_{m}+\omega_{eff})^{2} + \frac{A^{2} \omega_{0}^{4}/4}{(2\omega_{m} +\omega_{eff})^{2} - \omega_{0}^{2}} } \Big)\label{omegaeff2}
\eea
 We solve this equation iteratively in the driving amplitude $A$, which gives
\bea
\omega_{eff}^{2} &\approx & \omega_{0}^{2} + \frac{A^{2} \omega_{0}^{4}}{2 (\omega_{m}^{2} - 4 \omega_{0}^{2})} + \frac{25 A^{4} \omega_{0}^{8}}{32 \omega_{m}^{6}}.
\eea
Here, we kept the leading order in the inverse frequency $1/\omega_{m}$ for the fourth order term, which scales as $1/\omega_{m}^{6}$. We kept all orders in $1/\omega_{m}$ for the term that is second ordering $A$.


\section{Fourth order term of the Magnus expansion}\label{ME4}
 After expanding Eq. \ref{L4} to the order $k=3$, evaluating the integrals of the form of Eq. \ref{cint}, and collecting the terms that scale as $1/\omega_{m}^{6}$
 we obtain for $L_{eff}^{(4,6)}$:
   \begin{widetext}
 \bea
    L_{eff}^{(4,6)}
 &=& \frac{1}{12 \omega_{m}^{6}} \Big( \frac{1}{6} \frac{45}{64} [L_{dr,0}, [[  \text{ad}_{L_{0}}^{3} L_{dr,0}  ,L_{dr,0}],L_{dr,0}]]
   +\frac{1}{2} \frac{1}{32}  [L_{dr,0}, [[  \text{ad}_{L_{0}}^{2} L_{dr,0}, \text{ad}_{L_{0}} L_{dr,0}],L_{dr,0}]]\nonumber\\
 & &  + \frac{1}{2} \frac{7}{16}  [L_{dr,0}, [[  \text{ad}_{L_{0}} L_{dr,0}, \text{ad}_{L_{0}}^{2} L_{dr,0}],L_{dr,0}]]   
   - \frac{1}{6} \frac{9}{8}  [L_{dr,0}, [[  L_{dr,0}, \text{ad}_{L_{0}}^{3} L_{dr,0}],L_{dr,0}]]\nonumber\\    
 &&  + \frac{1}{2} \frac{27}{64} [ \text{ad}_{L_{0}} L_{dr,0}, [[  \text{ad}_{L_{0}}^{2} L_{dr,0}  ,L_{dr,0}],L_{dr,0}]]
   -  \frac{1}{2} \frac{15}{16}  [ \text{ad}_{L_{0}} L_{dr,0}, [[  L_{dr,0}, \text{ad}_{L_{0}}^{2} L_{dr,0}],L_{dr,0}]]\nonumber\\
 &&  + \frac{1}{2} \frac{21}{64} [ \text{ad}_{L_{0}}^{2} L_{dr,0}, [[  \text{ad}_{L_{0}} L_{dr,0}  ,L_{dr,0}],L_{dr,0}]]
   -  \frac{1}{2} \frac{33}{32}  [ \text{ad}_{L_{0}}^{2} L_{dr,0}, [[ L_{dr,0},  \text{ad}_{L_{0}} L_{dr,0}],L_{dr,0}]]\nonumber\\
&&   + \frac{1}{6} \frac{45}{64} [[L_{dr,0},  [ \text{ad}_{L_{0}}^{3} L_{dr,0}  ,L_{dr,0}]],L_{dr,0}]
   +\frac{1}{2}\frac{1}{32} [[L_{dr,0}, [   \text{ad}_{L_{0}}^{2} L_{dr,0} , \text{ad}_{L_{0}} L_{dr,0}  ]],L_{dr,0}]\nonumber\\
 & &  +\frac{1}{2}\frac{7}{16}  [[L_{dr,0}, [   \text{ad}_{L_{0}} L_{dr,0} , \text{ad}_{L_{0}}^{2} L_{dr,0}  ]],L_{dr,0}]
   -  \frac{1}{6}\frac{9}{8}  [[L_{dr,0}, [   L_{dr,0} , \text{ad}_{L_{0}}^{3} L_{dr,0}  ]],L_{dr,0}]\nonumber\\
 &&  + \frac{1}{2}\frac{27}{64} [[ \text{ad}_{L_{0}} L_{dr,0}, [   \text{ad}_{L_{0}}^{2} L_{dr,0}  , L_{dr,0} ]],L_{dr,0}]
   - \frac{1}{2}\frac{15}{16} [[ \text{ad}_{L_{0}} L_{dr,0}, [   L_{dr,0}, \text{ad}_{L_{0}}^{2} L_{dr,0}  ]],L_{dr,0}]\nonumber\\
 &&  + \frac{1}{2}\frac{21}{64} [[  \text{ad}_{L_{0}}^{2} L_{dr,0}, [    \text{ad}_{L_{0}} L_{dr,0}, L_{dr,0} ]],L_{dr,0}]
   -  \frac{1}{2}\frac{33}{32} [[  \text{ad}_{L_{0}}^{2} L_{dr,0}, [   L_{dr,0}, \text{ad}_{L_{0}} L_{dr,0}]]  ,L_{dr,0}]\nonumber\\
&&  - \frac{1}{2}\frac{33}{32} [[ \text{ad}_{L_{0}}^{2} L_{dr,0},L_{dr,0}] , [ \text{ad}_{L_{0}} L_{dr,0} ,L_{dr,0}]]
 + \frac{1}{2}\frac{1}{32} [[ L_{dr,0}, \text{ad}_{L_{0}}^{2} L_{dr,0}] , [ \text{ad}_{L_{0}} L_{dr,0},L_{dr,0}]]\nonumber\\
&&  - \frac{1}{2} \frac{15}{16} [[   \text{ad}_{L_{0}} L_{dr,0} ,L_{dr,0}] , [ \text{ad}_{L_{0}}^{2} L_{dr,0},L_{dr,0}]]
 + \frac{1}{2}\frac{7}{16} [[ L_{dr,0}, \text{ad}_{L_{0}} L_{dr,0}, [  \text{ad}_{L_{0}}^{2} L_{dr,0},L_{dr,0}]]\nonumber\\
&& + \frac{1}{2}\frac{21}{64} [[ \text{ad}_{L_{0}}^{2} L_{dr,0},L_{dr,0}] , [  \text{ad}_{L_{0}} L_{dr,0} ,L_{dr,0}]]
 + \frac{1}{2}\frac{7}{16} [[ L_{dr,0}, \text{ad}_{L_{0}}^{2} L_{dr,0} ] , [  \text{ad}_{L_{0}} L_{dr,0}  ,L_{dr,0}]]\nonumber\\
&&  +\frac{1}{2}\frac{27}{64} [[  \text{ad}_{L_{0}} L_{dr,0},L_{dr,0}] , [  \text{ad}_{L_{0}}^{2} L_{dr,0} ,L_{dr,0}]]
 + \frac{1}{2}\frac{1}{32} [[ L_{dr,0},\text{ad}_{L_{0}} L_{dr,0}], [  \text{ad}_{L_{0}}^{2} L_{dr,0}  ,L_{dr,0}]]\Big)
 \eea 
 
 \end{widetext}
 To simplify this expression, we first combine the terms that are related by commutation.
  Additionally, we use that 
 \bea
 &&  [ \text{ad}_{L_{0}} L_{dr,0}, [[  \text{ad}_{L_{0}}^{2} L_{dr,0}  ,L_{dr,0}],L_{dr,0}]]\\
  &&  + [ \text{ad}_{L_{0}}^{2} L_{dr,0}, [[  \text{ad}_{L_{0}} L_{dr,0}  ,L_{dr,0}],L_{dr,0}]]\\
  &=&   [[ \text{ad}_{L_{0}} L_{dr,0}, [   \text{ad}_{L_{0}}^{2} L_{dr,0}  , L_{dr,0} ]],L_{dr,0}]\\
    &&  + [[  \text{ad}_{L_{0}}^{2} L_{dr,0}, [    \text{ad}_{L_{0}} L_{dr,0}, L_{dr,0} ]],L_{dr,0}]
  \eea
and
 \bea
&& [ \text{ad}_{L_{0}} L_{dr,0}, [   \text{ad}_{L_{0}}^{2} L_{dr,0}  , L_{dr,0} ]]\\
&=& [ \text{ad}_{L_{0}}^{2} L_{dr,0}, [   \text{ad}_{L_{0}} L_{dr,0}  , L_{dr,0} ]]\\
&& - [  L_{dr,0}, [ \text{ad}_{L_{0}} L_{dr,0},  \text{ad}_{L_{0}}^{2} L_{dr,0}   ]] 
 \eea
 With these identities, we simplify the expression to the form given in Eq. \ref{L46}.

%


%

\end{document}